\documentclass[runningheads]{llncs}

\usepackage{xcolor}

\usepackage{amsmath}
\usepackage{amssymb}

\usepackage{booktabs}
\usepackage{pgfplots}
\usepackage[hidelinks]{hyperref}
\usepackage{listings}
\usepackage{multirow}
\usepackage{comment}
\usepackage{pgfplots}
\usepackage{array}
\pgfplotsset{compat=newest}
\usepackage{float}
\usepackage[ruled,noline,noend,linesnumbered]{algorithm2e}
\usepackage[nounderscore]{syntax}

\SetCommentSty{mycommfont}
\SetKw{KwAnd}{and}
\SetKw{KwNot}{not}
\SetKw{KwBreak}{break}
\SetKw{KwInfl}{in}
\SetKwProg{Pn}{}{:}{\KwRet}

\usepackage{afterpage}

\usepackage{pgf}
\usepackage{tikz}
\usetikzlibrary{arrows,shapes,automata,backgrounds}

\let\emptyset\varnothing

\definecolor{mygray}{gray}{0.6}

\def\plevent{\textsf{\small event}}
\def\plstate{\textsf{\small state}}
\def\pldynamic{\textsf{\small dynamic}}

\def\plstart{\textsf{\small start}}
\def\plend{\textsf{\small end}}

\def\plbefore{\textsf{\small before}}
\def\plmeets{\textsf{\small meets}}
\def\ploverlaps{\textsf{\small overlaps}}
\def\plfinishes{\textsf{\small finishes}}
\def\plcontains{\textsf{\small contains}}

\def\plstarts{\textsf{\small starts}}
\def\plequals{\textsf{\small equals}}

\newcommand{\mt}[1]{{\mathit{#1}}}

\DeclareMathSymbol{\mlq}{\mathord}{operators}{``}
\DeclareMathSymbol{\mrq}{\mathord}{operators}{`'}

\newenvironment{mysplit}%
  {\arraycolsep 0pt \begin{array}{l}}%
  {\end{array}}

\newcommand\arraybslash{\let\\\@arraycr}
\newcounter{Table}

\usepackage{xcolor}

\makeatletter
\newcommand*\bigcdot{\mathpalette\bigcdot@{.5}}
\newcommand*\bigcdot@[2]{\mathbin{\vcenter{\hbox{\scalebox{#2}{$\m@th#1\bullet$}}}}}
\makeatother

\begin{document}

\title{Representation and Processing of Instantaneous and Durative Temporal Phenomena\thanks{This work has been funded by the Engineering and Physical Sciences Research Council (EPSRC) Centre for Doctoral Training in Distributed Algorithms at the University of Liverpool, and Denbridge Marine Limited, United Kingdom.}}

\author{Manolis Pitsikalis\orcidID{0000-0003-2959-2022} \and
Alexei Lisitsa\orcidID{0000-0002-3820-643X} \and
Shan Luo\orcidID{0000-0003-4760-0372}}

\institute{Department of Computer Science, University of Liverpool\\
\email{\{e.pitsikalis,a.lisitsa,shan.luo\}@liverpool.ac.uk}}

\authorrunning{M. Pitsikalis et al.} 

\maketitle

\begin{abstract} 
Event definitions in Complex Event Processing systems are constrained by the expressiveness of each system's language. Some systems allow the definition of instantaneous complex events, while others allow the definition of durative complex events. While there are exceptions that offer both options, they often lack of intervals relations such as those specified by the Allen's interval algebra. In this paper, we propose a new logic based temporal phenomena definition language, specifically tailored for Complex Event Processing, that allows the representation of both instantaneous and durative phenomena and the temporal relations between them. Moreover, we demonstrate the expressiveness of our proposed language by employing a maritime use case where we define maritime events of interest. Finally, we analyse the execution semantics of our proposed language for stream processing and introduce the `Phenesthe' implementation prototype.
\keywords{Event definition language, temporal logic, stream processing, event recognition}
\end{abstract}

\section{Introduction}
There are numerous event description languages, each with its own formal description and expressiveness. Event description languages allow the representation and the specification of temporal phenomena. They have been used widely in, among others, complex event processing and recognition~\cite{Cugola_Margara_2010,Anicic_2010,Artikis_2015,Beck_2018},  system analysis and verification~\cite{chen1991,bellini2000}. In the case of Complex Event Processing, which is the focus of this paper, users with expert knowledge, or machine learning algorithms, provide definitions of events of interest i.e., complex events that are represented in an event definition language. The Complex Event Processing system accepts as input a single or multiple streams of low level events, such as the timestamped transmitted values of a sensor and by continuously applying temporal queries that involve the provided event definitions, it will produce a stream of complex events associated with some temporal information.

However, the set of events that can be represented in the language of a Complex Event Processing system is constrained  by the expressiveness of its event definition language. For example, languages with a point-based temporal model associate facts to instants of time while languages with an interval-based temporal model associate facts with intervals~\cite{Chomicki_1994}, therefore the representation of durative and instantaneous entities in each case respectively is sometimes impossible or not straightforward. Event Calculus~\cite{kowalski86} approaches~\cite{Artikis_2015,chittaro1996,khan2019} allow the representation of both instantaneous and durative entities, however they lack of interval relations such as those specified by the Allen's interval algebra~\cite{Allen_1983}. 

In this paper, we formally present a logic based language for Complex Event Processing that allows the description of both instantaneous and durative temporal phenomena and the relations between them.  Moreover, we present the execution semantics for stream processing and demonstrate the expressiveness of the language by employing a maritime use-case scenario where the goal is to describe maritime activities of interest. Thus, the contributions of this paper are the following:
\begin{itemize}
\item a formally described language for the representation of both instantaneous and durative temporal phenomena and their relations,
\item the execution semantics for stream processing along with an implementation prototype,
\item a demonstration of the expressiveness of the language in a maritime use case.
\end{itemize}

The paper is organised as follows, in Section~\ref{sec:language} we present the syntax, the grammar and the semantics of our language. Next, in  Section~\ref{sec:maritime} we demonstrate its expressiveness by formalising a set of maritime activities. In Section~\ref{sec:exsemantics} we present the execution semantics for stream processing, while in Section \ref{sec:related} we compare our language with relevant works. Finally, in Section~\ref{sec:summary} we summarise our approach and discuss our future directions. 
\section{Language}
\label{sec:language}
The key components of our language are events, states and dynamic temporal phenomena. In what follows, `temporal phenomena' includes all of the three aforementioned entities. Events are true at instants of time, while states and dynamic temporal phenomena hold on intervals. Events are defined in terms of logical operations between instantaneous temporal phenomena, states are defined using the operators of maximal range, temporal union, temporal intersection and temporal complement; finally, dynamic temporal phenomena are defined in terms of temporal relations that involve the basic seven of Allen's interval algebra~\cite{Allen_1983}.  In this Section, we present the syntax, the semantics  and the grammar of our language. 

\subsection{Syntax}
Formally, our language is described by the triplet $\langle \mathcal{P},L,\Phi \rangle$, where
\begin{itemize}
\item $\mathcal{P}$ is a set of Predicates (atomic formulae), that may be of three types, event, state or dynamic temporal phenomenon predicates;
\item $L$ is a set  defined by the union of the set of logical connectives $\{\wedge,\vee,\neg\}$, the set of temporal operators, $\{\rightarrowtail, \sqcup,\sqcap,\setminus\}$, the set of temporal relations $\{\plbefore$, $\plmeets$, $\ploverlaps$, $\plfinishes$, $\plstarts$, $\plequals$, $\plcontains\}$ and finally the set of the $\{\plstart,\plend\}$ operators.
\item $\Phi$ is the set of formulae defined by the union of the formulae sets $\Phi^{\bigcdot}$, $\Phi^-$ and $\Phi^=$, that we will present below.
\end{itemize}
Formulae of $\Phi^{\bigcdot}$ describe instantaneous temporal phenomena, formulae of $\Phi^-$ describe durative temporal phenomena that hold (are true) in disjoint maximal intervals, finally formulae of $\Phi^=$ describe durative temporal phenomena that may hold in non-disjoint intervals. Atomic formulae of $\Phi^\square$ where $\square \in\{\bigcdot,-,=\}$ are denoted as $\Phi^\square_o$.
Terms are defined as follows:
\begin{itemize}
\item Each variable is a term.
\item Each constant is a term.
\end{itemize}
Events, states and dynamic temporal phenomena are expressed as n-ary predicate symbols, of the corresponding type (event, state or dynamic temporal phenomenon), $\mathit{P(a_1,...,a_n)}$, where $P$ is the associated name and $a_1,...,a_n$ are terms corresponding to atemporal properties. Moreover, we assume that the set of predicate symbols includes those with atemporal and fixed semantics, such as arithmetic comparison operators etc., however for simplification reasons in what follows we omit their presentation. 
The set of formulae $\Phi^{\bigcdot}$ is defined as follows:
\begin{itemize}
\item (Event predicate) If P is an n-ary event predicate and $a_1,...,a_n$ are terms then $P(a_1,...,a_n)$ is a formula of $\Phi^{\bigcdot}$ .
\item (Negation) If $\phi$ is a formula of $\Phi^{\bigcdot}$  then $\neg\phi$ is a formula of $\Phi^{\bigcdot}$ .
\item (Conjunction/disjunction) If $\phi$ and $\psi$ are formulae of $\Phi^{\bigcdot}$  then $\phi\ op\ \psi$, where $op$ is the conjunction ($\wedge$) or disjunction ($\vee $) connective, is a formula of $\Phi^{\bigcdot}$.
\item (start/end) If $\phi$ is a formula of $\Phi^-$
then  $\plstart(\phi)$ and $\plend(\phi)$ are formulae of $\Phi^{\bigcdot}$.
\end{itemize}
Following, we define the set of formulae $\Phi^-$:
\begin{itemize}
\item (State predicate) If P is an n-ary state predicate and $a_1,...,a_n$ are terms then $P(a_1,...,a_n)$ is a formula of $\Phi^-$.
\item (Maximal range) If $\phi$ and $\psi$ are formulae of $\Phi^{\bigcdot}$ then $\phi\rightarrowtail\psi$ is a formula of $\Phi^-$. 
\item (Temporal union, intersection \& complement) If $\phi$ and $\psi$ are formulae of $\Phi^-$ then $\phi\ \square\ \psi$, where $\square\in\{\sqcup,\sqcap, \setminus\}$, is a formula of $\Phi^-$. The temporal operators $\sqcup,\sqcap$ and $\setminus$ correspond to temporal union, intersection and complement respectively. 
\end{itemize}
Finally we define the set of formulae $\Phi^=$ as follows:
\begin{itemize}
\item(Dynamic temporal phenomenon predicate) If P is an n-ary dynamic temporal phenomenon predicate and $a_1,...,a_n$ are terms then $P(a_1,...,a_n)$ is a formula of $\Phi^=$.
\item(Temporal relation)
\begin{itemize}
\item If $\phi$ and $\psi$ are formulae of $\Phi^-\cup\ \Phi^=$ then $\phi\ \mt{tr}\ \psi$, where $\mt{tr} \in \{\plmeets,$ $\ploverlaps,$ $\plequals\}$, is a formula of $\Phi^=$.
\item If $\phi$ is a formula of $\Phi^{\bigcdot}\cup\Phi^-\cup\Phi^=$ and $\psi$ is a formula of $\Phi^-\cup\Phi^=$ then $\phi\ \mt{tr}\ \psi$, where $\mt{tr} \in \{\plfinishes, \plstarts\}$, is a formula of $\Phi^=$. 
\item If $\phi$ is a formula of $\Phi^-\cup\Phi^=$ and $\psi$ is a formula of $\Phi^{\bigcdot}\cup\Phi^-\cup\Phi^=$ then $\phi\ \plcontains\ \psi$ is a formula of $\Phi^=$. 
\item If $\phi$ is a formula of $\Phi^{\bigcdot}\cup\Phi^-\cup\Phi^=$ and $\psi$ is a formula of $\Phi^{\bigcdot}\cup\Phi^-\cup\Phi^=$  then $\phi\ \plbefore\ \psi$ is a formula of $\Phi^=$. 
\end{itemize}
\end{itemize}
\subsection{Grammar}
\label{sec:grammar}
In addition to the elements described above, our language includes temporal phenomena definitions
 that specify new event, state and dynamic temporal phenomena predicates. The production rule-set~\eqref{eq:ebnf-all} presents the complete grammar of the language in the Extended Backus-Naur form (EBNF). 
 {
    \scriptsize
\begin{align}
\begin{mysplit}
\label{eq:ebnf-all}
\langle \mt{event} \rangle\ ::=\ \mt{eventName}(...);\\
\langle \mt{state} \rangle\ ::=\ \mt{stateName}(...);\\
\langle \mt{dynamic} \rangle\ ::=\ \mt{dynamicPhenomenonName}(...);\\[2.5pt]
\langle \mt{temporalExpression} \rangle\ ::=\ \langle \mt{instantExpression} \rangle\ |\ \langle \mt{intervalExpression} \rangle;\\[2.5pt]
\langle \mt{instantExpression} \rangle\ ::=\ \mlq(\mrq \langle \mt{instantExpresstion} \rangle \mlq)\mrq\ |\ \neg\ \langle \mt{instantExpression} \rangle\\
 \qquad\qquad\qquad\qquad\qquad\qquad|\  \langle \mt{instantExpression} \rangle\  ( \mlq\wedge\mrq\ |\ \mlq\vee\mrq )\ \langle \mt{instantExpression} \rangle \\ 
 \qquad\qquad\qquad\qquad\qquad\qquad|\ \langle \mt{startEndOp} \rangle\ |\  \langle \mt{event} \rangle;\\[2.5pt]
 \langle \mt{intervalExpression} \rangle\ ::=\ \langle \mt{intervalOperation} \rangle\ |\ \langle \mt{intervalRelation} \rangle;\\[2.5pt]
 \langle \mt{intervalOperation} \rangle\ ::=\ \langle \mt{intervalOperation} \rangle\  ( \mlq\sqcup\mrq\ |\ \mlq\sqcap\mrq\ |\ \mlq\setminus\mrq)\ \langle \mt{intervalOperation} \rangle\\
 \qquad\qquad\qquad\qquad\qquad\qquad|\ \langle \mt{instantExpression} \rangle\ \mlq\rightarrowtail\mrq\ \langle \mt{instantExpression} \rangle\\
 \qquad\qquad\qquad\qquad\qquad\qquad|\ \ \mlq(\mrq \langle \mt{intervalOperation} \rangle\mlq)\mrq\ |\  \langle state \rangle;\\[2.5pt]
 \langle \mt{intervalRelation} \rangle\ ::=\ \langle \mt{temporalExpression} \rangle\  \mlq\texttt{before}\mrq\  \langle \mt{temporalExpression} \rangle \\ 
 \qquad\qquad\qquad\qquad\qquad\quad|\ \langle \mt{intervalExpression} \rangle\  \mlq\texttt{overlaps}\mrq\  \langle \mt{intervalExpression} \rangle \\ 
 \qquad\qquad\qquad\qquad\qquad\quad|\ \langle \mt{intervalExpression} \rangle\  \mlq\texttt{meets}\mrq\  \langle \mt{intervalExpression} \rangle \\ 
  \qquad\qquad\qquad\qquad\qquad\quad|\ \langle \mt{temporalExpression} \rangle\  \mlq\texttt{finishes}\mrq\  \langle \mt{intervalExpression} \rangle \\ 
   \qquad\qquad\qquad\qquad\qquad\quad|\ \langle \mt{temporalExpression} \rangle\  \mlq\texttt{starts}\mrq\  \langle \mt{intervalExpression} \rangle \\ 
    \qquad\qquad\qquad\qquad\qquad\quad|\ \langle \mt{intervalExpression} \rangle\  \mlq\texttt{contains}\mrq\  \langle \mt{temporalExpression} \rangle \\ 
     \qquad\qquad\qquad\qquad\qquad\quad|\ \langle \mt{intervalExpression} \rangle\  \mlq\texttt{equals}\mrq\  \langle \mt{intervalExpression} \rangle \\ 
     \qquad\qquad\qquad\qquad\qquad\quad|\ \mlq(\mrq \langle \mt{intervalRelation} \rangle  \mlq)\mrq\ |\ \langle \mt{dynamic} \rangle;\\[2.5pt]
\langle \mt{startEndOp} \rangle\ ::=\ (\mlq \texttt{start} \mrq\ |\ \mlq \texttt{end} \mrq)  \mlq(\mrq\langle \mt{intervalOperation} \rangle\mlq)\mrq;\\[2.5pt]
\langle \mt{definitions} \rangle\ ::=\  \langle \mt{eventDefinition} \rangle\ |\ \langle \mt{stateDefinition} \rangle\\
\qquad\qquad\qquad\qquad\quad |\  \langle \mt{dynamicDefinition}\rangle;\\[2.5pt]
 \langle \mt{eventDefinition} \rangle\ ::=\ \mlq\texttt{event} \mrq\ \langle event \rangle\mlq:\mrq\ \langle \mt{instantExpression} \rangle\mlq.\mrq;\\
\langle \mt{stateDefinition} \rangle ::= \mlq\texttt{state}\mrq\ \langle state\rangle \mlq:\mrq \langle \mt{intervalOperation} \rangle\mlq.\mrq;\\
\langle \mt{dynamicDefinition}\rangle ::= \mlq\texttt{dynamic}\mrq\ \langle dynamic \rangle \mlq:\mrq\langle \mt{intervalRelation}\rangle\mlq.\mrq;
\end{mysplit}
\end{align}
}

Phenomena definitions are specified by the production rule $\langle \textit{definitions} \rangle$ of the grammar. Event predicates are defined using expressions on instants of time, expressed via formulae of $\Phi^{\bigcdot}$, while state predicates are defined in terms of formulae of $\Phi^-$. Finally, dynamic temporal phenomena predicates are defined in terms of temporal relations on intervals and instants, therefore specified via formulae of  $\Phi^=$. Note, that the set of definitions allowed in the language is subject to an additional constraint: no cyclic dependencies in the definitions of temporal phenomena are allowed.  A temporal phenomenon $A$ depends from phenomenon $B$ if $B$ is a sub-formula of the definition of $A$.

\subsection{Semantics}
\label{sec:semantics}
We assume time is represented by an infinite non empty set $T=\mathbb{Z}^+_0$ of non-negative integers ordered via the `$<$' relation, formally  $\mathcal{T}=\langle T, < \rangle$. 
For the formulae sets $\Phi^{\bigcdot}, \Phi^-$ and $\Phi^=$ we define the model $\mathcal{M}=\langle T, I, <,V^{\bigcdot}, V^-, V^= \rangle$ where $V^.:\Phi_o^.\rightarrow 2^T$,  $V^-:\Phi_o^-\rightarrow 2^I$, $V^=:\Phi_o^=\rightarrow 2^I$ are valuations of atomic formulae and $I=\{ [ts,te] :ts<te\ \text{and}\ ts,te \in T\}\cup\{[ts,\infty): ts \in T\}$ is the set of the accepted time intervals of $T$. 
Given a model $\mathcal{M}$, the validity of a formula $\phi\in\Phi^{\bigcdot}$ at a timepoint $t\in T$ (in symbols $\mathcal{M},t\models\phi$) is determined by the rules below:
\begin{itemize}
\item $\mathcal{M},t\models P(a_1,...,a_n)$ where $P$ is an n-ary event predicate symbol iff $t\in$ $ V^.(P(a_1,...,a_n))$;
\item $\mathcal{M},t\models \neg \phi$ where $\phi\in\Phi^{\bigcdot}$ iff $\mathcal{M},t\not\models\phi$;
\item $\mathcal{M},t\models \phi \wedge \psi$ where $\phi,\psi\in\Phi^{\bigcdot}$ iff $\mathcal{M},t\models \phi$ and $\mathcal{M},t\models \psi$;
\item $\mathcal{M},t\models \phi \vee \psi$ where $\phi,\psi\in\Phi^{\bigcdot}$ iff $\mathcal{M},t\models \phi$ or $\mathcal{M},t\models \psi$;
\item $\mathcal{M},t\models \plstart(\phi)$ where $\phi\in\Phi^-$ iff $\exists te \in T$ and $\mathcal{M},[t,te] \models \phi$, where $\mathcal{M},[t,te] \models \phi$ denotes the validity of formula $\phi\in\Phi^-$ at an interval $[t,te]$ as defined below;
\item $\mathcal{M},t\models \plend(\phi)$ where $\phi\in\Phi^-$ iff $\exists ts \in T$ and $\mathcal{M},[ts,t] \models \phi$;
\end{itemize}
Given a model $\mathcal{M}$, the validity of a formula $\phi\in\Phi^-$ at a time interval $[ts,te]\in I$ (in symbols $\mathcal{M},[ts,te]\models\phi$) is defined as follows:
\begin{itemize}
\item $\mathcal{M},[ts,te]\models P(a_1,...,a_n)$ where $P$ is an n-ary state predicate symbol iff $[ts,te]\in $  $V^-(P(a_1,...,a_n))$;
\item $\mathcal{M},[ts,te]\models \phi \rightarrowtail \psi$ where $\phi,\psi\in\Phi^{\bigcdot}$ iff:
\begin{enumerate}
\item $ts \in T$ and $\mathcal{M},ts\models \phi$,
\item $te \in (ts,\infty)\subset T$ and $\mathcal{M},te\models \psi \wedge \neg \phi$,
\item $\forall ts' \in [0,ts) \subset T,\exists te'\in(ts',ts)$ where $\mathcal{M},ts'\models\phi\ \text{and}\ \mathcal{M},te'\models \psi \wedge \neg \phi$,
\item and finally,  $\nexists te'' \in (ts,te)\subset T $  where $\mathcal{M},te''\models\psi\wedge\neg\phi$.
\end{enumerate} 
Essentially, $\phi \rightarrowtail \psi$ holds for the disjoint maximal intervals that start at the earliest instant $ts$ where $\phi$ is true (conditions 1,3) and end at the earliest instant $te,  te>ts$ (condition 4)  where $\psi$ is true and $\phi$ is false (condition 2,4).
\item $\mathcal{M},[ts,\infty) \models \phi \rightarrowtail \psi$ where $\phi,\psi\in\Phi^{\bigcdot}$ iff conditions (1) and (3) from above hold and $\nexists te \in (ts,\infty)\subset T$ such that $\mathcal{M},te\models \psi \wedge \neg \phi$. Therefore a formula  $\phi \rightarrowtail \psi$ may hold indefinitely if there does not exist an appropriate instant after $ts$ at which $\phi\wedge\neg\psi$ is satisfied. This effectively implements the commonsense law of inertia~\cite{Mueller_2015}. For simplification reasons in the semantics below we omit intervals open at the right to infinity since they can be treated in a similar manner.
\item $\mathcal{M},[ts,te]\models \phi\ \sqcup\ \psi$ where $\phi,\psi\in\Phi^-$ iff one of the cases below holds:
\begin{itemize}
\item $\exists$ a sequence of length $k > 1$ of intervals $i_1,...,i_k \in I$ where $i_k=[ts_k,te_k]$, $ts=ts_1$ and $te=te_k$  s.t.:
\begin{enumerate}
\item $\forall\alpha\in [1,k-1]$: $te_\alpha\in i_{\alpha+1}$ and $ts_\alpha < ts_{\alpha+1}$ and $te_\alpha < te_{\alpha+1}$  ,
\item $\forall\beta\in[1,k]$:  $\mathcal{M},[ts_\beta,te_\beta]\models\phi$ or $\mathcal{M},[ts_\beta,te_\beta]\models\psi$, and
\item $\nexists i_\gamma=[ts_\gamma,te_\gamma] \in I$ where  $\mathcal{M},[ts_\gamma,te_\gamma]\models\phi$ or $\mathcal{M},[ts_\gamma,te_\gamma]\models\psi$ and $ts_1 \in i_\gamma$ or $te_k \in i_\gamma$
\end{enumerate}
\item $\mathcal{M},[ts,te]\models\phi$ or $\mathcal{M},[ts,te]\models\psi$ and $\nexists i_\gamma=[ts_\gamma,te_\gamma] \in I$ where $\mathcal{M},[ts_\gamma,te_\gamma]\models\phi$ or $\mathcal{M},[ts_\gamma,te_\gamma]\models\psi$  and $ts \in i_\gamma$ or $te \in  i_\gamma$.
\end{itemize}
For a sequence of intervals, conditions (1-2) ensure that intervals, at which $\phi$ or $\psi$ are valid,  overlap or touch will coalesce, while condition (3) ensures that the resulting interval is maximal. In the case of a single interval, the conditions ensure that at the interval $[ts,te]$  $\phi$ or $\psi$ is valid, and that $[ts,te]$ is maximal. In simple terms, the temporal union $\phi\ \sqcup\ \psi$ holds for the intervals where at least one of $\phi$ or $\psi$ hold. The above definition of temporal union follows the definitions of temporal coalescing presented in~\cite{bohlen1998,Dohr2018}.
\item $\mathcal{M},[ts,te]\models \phi\ \setminus\ \psi$ where $\phi,\psi\in\Phi^-$ iff $\exists  [ts',te'] \in I$ where $\mathcal{M},[ts',te']\models \phi$, $[ts,te]\subseteq [ts',te']$ (i.e. $[ts,te]$ subinterval of $[ts',te']$), $\forall [ts_\psi,te_\psi]\in I$ where $\mathcal{M},[ts_\psi,te_\psi]\models \psi$,  $[ts,te]\cap [ts_\psi,te_\psi]=\emptyset$ and finally $[ts,te]$ is maximal. In plain language, the temporal difference of formulae $\phi,\psi$ holds for the maximal subintervals of the intervals at which $\phi$ holds but $\psi$ doesn't hold.  
\item $\mathcal{M},[ts,te] \models \phi\ \sqcap\ \psi$ where $\phi,\psi\in\Phi^-$  iff $\exists  [ts_\phi,te_\phi], [ts_\psi, te_\psi] \in I$ where $\mathcal{M},[ts_\phi,te_\phi]\models \phi$, $\mathcal{M},[ts_\psi,te_\psi]\models \psi$ and $\exists [ts,te]\in I$ where $[ts,te] \subseteq [ts_\phi,te_\phi]$, $[ts,te] \subseteq [ts_\psi,te_\psi]$ and $[ts,te]$ is maximal. In other words, the temporal intersection of two formulae of $\Phi^-$ holds for the intervals at which both formulae hold.
\end{itemize}
Given a model $\mathcal{M}$, the validity of a formula $\phi\in\Phi^=$ at a time interval $[ts,te]\in I$ (in symbols $\mathcal{M},[ts,te]\models\phi$) is defined as follows:
\begin{itemize}
\item $\mathcal{M},[ts,te]\models P(a_1,...,a_n)$ where $P$ is n-ary dynamic temporal phenomenon predicate symbol iff $[ts,te]\in V^=(P(a_1,...,a_n))$;
\item  $\mathcal{M},[ts,te]\models \phi\ \plbefore\ \psi$ iff:
\begin{itemize}
%
%
\item (interval - interval) for $\phi,\psi \in \Phi^-\cup\Phi^=$, $\exists a, b \in T$, $a<b$, $\mathcal{M},[ts,a]\models \phi$, $\mathcal{M},[b,te]\models \psi$ and all of the conditions below hold:
\begin{itemize}
\item $\nexists [ts',a']\in I$ where $\mathcal{M},[ts',a']\models \phi$ and $a < a'< b$,
\item $\nexists [b',te']\in I$ where $\mathcal{M},[b',te']\models \psi$ and $a <b' < b$.
\end{itemize}
%
%
\item (instant - interval) for $\phi \in \Phi^{\bigcdot}$, $\psi \in \Phi^-\cup\Phi^=$, $\mathcal{M},ts \models \phi$, $\exists a \in T, a>ts$, $\mathcal{M},[a,te] \models \psi$ and all the conditions below hold:
\begin{itemize}
\item $\nexists ts' \in T$ where $\mathcal{M},ts'\models \phi$ and $ts < ts'< a$,
\item $\nexists[a',te']\in I$ where $\mathcal{M},[a',te']\models \psi$ and $ts <a' < a$.
\end{itemize}
%
%
\item (interval - instant) for $\phi \in \Phi^-\cup\Phi^=$ and $\psi \in \Phi^{\bigcdot}$, $\exists a \in T$, $a<te$, $\mathcal{M},[ts,a] \models \phi$, $\mathcal{M},te \models \psi$ and all the conditions below hold:
\begin{itemize}
\item  $\nexists [ts',a']\in I$ where $\mathcal{M},[ts',a']\models \phi$ and $a < a'< te$,
\item $\nexists te' \in T$ where $\mathcal{M},te'\models \psi$ and $a < te' < te$.
\end{itemize}
%
%
\item (instant - instant) for $\phi \in \Phi^{\bigcdot}$ and $\psi \in \Phi^{\bigcdot}$, $\mathcal{M},ts \models \phi$, $\mathcal{M},te \models \psi$ and all the conditions below hold:
\begin{itemize}
\item $\nexists ts'$ where $\mathcal{M},ts'\models \phi$ and $ts < ts'< te$,
\item $\nexists te'$ where $\mathcal{M},te'\models \psi$ and $ts < te' < te$.
\end{itemize}
\end{itemize}
In our approach the `before' relation holds only for intervals where the pair of instants or intervals at which the participating formulae are true or hold are contiguous. For example for the intervals $[1,2]$,$[1,3]$ and $[5,6]$ only $[1,3]$ is $\plbefore$ $[5,6]$. 
\item $\mathcal{M},[ts,te]\models \phi\ \plmeets\ \psi$ iff $\phi,\psi \in \Phi^-\cup\Phi^=$, $\exists a\in T$, $\mathcal{M},[ts,a]\models \phi$ and $\mathcal{M},[a,te]\models \psi$.
 \item $\mathcal{M},[ts,te]\models \phi\ \ploverlaps\ \psi$ iff $\phi,\psi \in \Phi^-\cup\Phi^=$, $\exists a, b \in T$, $ts < b < a <te$, $\mathcal{M},[ts,a]\models \phi$ and $\mathcal{M},[b,te]\models \psi$.
\item $\mathcal{M},[ts,te]\models \phi\ \plfinishes\ \psi$ iff:
\begin{itemize}
%
%
\item (interval - interval) for $\phi,\psi \in \Phi^-\cup\Phi^=$, $\exists a \in T$, $ts<a$, $\mathcal{M},[a,te]\models \phi$ and $\mathcal{M},[ts,te]\models \psi$,
%
%
 \item (instant - interval) for $\phi \in \Phi^{\bigcdot}$ and $\psi \in \Phi^-\cup\Phi^=$, $\mathcal{M},te \models \phi$ and $\mathcal{M},[ts,te]\models \psi$.
\end{itemize}
\item $\mathcal{M},[ts,te] \models\phi\ \plstarts\ \psi$ iff:
\begin{itemize} 
%
%
\item (interval - interval) for $\phi,\psi \in \Phi^-\cup\Phi^=$, $\exists a \in T$, $a<te$, $\mathcal{M},[ts,a]\models \phi$ and $\mathcal{M},[ts,te]\models \psi$,
%
%
 \item (instant - interval) for $\phi \in \Phi^{\bigcdot}$, $\psi \in \Phi^-\cup\Phi^=$, $\mathcal{M},ts \models \phi$ and $\mathcal{M},[ts,te]\models \psi$.
\end{itemize}
%
%
\item $\mathcal{M},[ts,te]\models \phi\ \plequals\ \psi$ iff  for $\phi,\psi \in \Phi^-\cup\Phi^=$, $\mathcal{M},[ts,te]\models \phi$  and $\mathcal{M},[ts,te]\models \psi$.
%
%
\item $\mathcal{M},[ts,te]\models \phi\ \plcontains\ \psi$ iff:
  \begin{itemize}
   \item (interval - interval) for $\phi,\psi \in \Phi^-\cup\Phi^=$, $\mathcal{M},[ts,te]\models \phi$, $\exists [a,b]\in I$, $ts< a< b<te$ and  $\mathcal{M},[a,b]\models \psi$,
%
%
 \item (interval - instant) for $\phi \in \Phi^-\cup\Phi^=$, $\psi \in \Phi^{\bigcdot}$, $\mathcal{M},[ts,te]\models \phi$, $\exists a \in T$, $ts< a <te$ and $\mathcal{M}, a \models \psi$.
\end{itemize}
\end{itemize}
\section{Maritime use case examples}
\label{sec:maritime}
Maritime situational awareness (MSA) is of major importance for environmental  and safety reasons. Maritime monitoring systems contribute in MSA by allowing the detection of possibly dangerous, or unlawful activities while also monitoring normal activities. Typically, maritime monitoring systems---as in the case of this paper---use data  from the Automatic Identification System (AIS) a system that allows the transmission of timestamped positional and vessel identity data from transceivers on vessels. Additionally, they may use contextual static information such as areas of interest or historical vessel information. Logic based approaches have already been used for Maritime Monitoring. Works such as~\cite{Pitsikalis_19,Roy2010} demonstrate that maritime activities can be expressed as patterns written in some form of an event definition language and be efficiently recognised. However, as mentioned earlier the definitions of temporal phenomena are constrained by the expressiveness of each language. In this section, we present definitions of maritime situations of interest written in our temporal phenomena definition language. 
\subsection{Stopped vessels}
A vessel is stopped when its speed is between a range  e.g., $0-0.5$ knots. Consider the definition for stopped vessels below:
\begin{align}
\begin{mysplit}
\plevent\ \mt{stop\_start(Vessel)} :\\
\quad  \mt{ais(Vessel, Speed, ...)} \wedge \mt{Speed} \leq 0.5\ .\\[4pt]
\plevent\ \mt{stop\_end(Vessel)} :\\
\quad  \mt{ais(Vessel, Speed, ...)} \wedge \mt{Speed} > 0.5\ .\\
\plstate\ \mt{stopped(Vessel)} :\\
\quad \mt{stop\_start(Vessel)}\rightarrowtail\mt{stop\_end(Vessel)}.
\end{mysplit}
\end{align}
$\mt{ais(Vessel, Speed, ...)}$ is an input event that contains the vessel id (Maritime Mobile Service Identity), its speed, its longitude and latitude and other AIS  information, while $\mt{stop\_start}$ and $\mt{stop\_end}$ are events that happen when an AIS message is received for a $\mt{Vessel}$ and its speed is either $\leq 0.5$ or $>0.5$ respectively. $\mt{stopped/1}$ is a state defined using the maximal range operator ($\rightarrowtail$) between the events $\mt{stop\_start}$ and $\mt{stop\_end}$, that holds true for the maximal intervals a vessel's speed is continuously $\leq$ 0.5 knots. Note that the $\mt{stop\_start}$ and $\mt{stop\_end}$ events are optional, since their definition could be integrated in the definition of the $\mt{stopped}$ state directly.

\subsection{Moored vessel} A vessel is considered moored when it is stopped near a port. The knowledge of the times and the locations vessels are moored is especially important in historical track analyses, law enforcement, etc. Below we provide the definition for moored vessels:
\begin{align} 
\begin{mysplit}
\label{eq:moored}
\plstate\ \mt{moored(Vessel,Port)}:\\
\quad  \mt{stopped(Vessel)}\ \sqcap \mt{in\_port(Vessel,Port)}.\\
\end{mysplit}
\end{align}
$\mt{moored/2}$ is a state defined using the interval operator of intersection between the $\mt{stopped/2}$ and the $\mt{in\_port(Vessel, Port)}$ states where the latter is an input state that holds when a $\mt{Vessel}$ is inside a $Port$. Therefore, according to the definition $\eqref{eq:moored}$, $\mt{moored}$ holds for the maximal sub-intervals of the intervals at which a vessel is stopped and inside a port (recall the semantics of the intersection operator as defined in Section~\ref{sec:semantics}).
\subsection{Fishing trips}
Monitoring of the fishing areas a fishing vessel has been through is important for sustainability and safety reasons. A fishing trip can be described as a series of certain maritime activities that are arranged over a long period of time as follows. Fishing vessels leave from a port, then they are under-way, they start a fishing activity in a fishing area, and then they return to the same or another port. A definition of the fishing trip temporal phenomenon is the following:
\begin{align}
\begin{mysplit}
\label{eq:fishtrip}
\pldynamic\ \mt{fishing\_trip(Vessel, PortA, AreaID, PortB)} :\\
\quad \mt{\plend(moored(Vessel,PortA))}\wedge\ type(Vessel, Fishing)\ \plbefore\\
\quad (\mt{underway(Vessel)}\ \plcontains\ \mt{in\_fishing\_area(Vessel,AreaID))}\\
\quad\ \plbefore\ \mt{\plstart(moored(Vessel, PortB))}.\\
\end{mysplit}
\end{align}
$\mt{in\_fishing\_area(Vessel,AreaID)}$ is an input state that  holds for the intervals a $\mt{Vessel}$ is within a fishing area with id $\mt{AreaID}$.  $\mt{underway}$ is a user defined state that holds for the intervals a vessel's speed is greater than 2.7 knots. We omit the presentation of the definition of under-way due to space limitations. Therefore, $\mt{fishing\_trip/4}$ as defined in rule~\eqref{eq:fishtrip}, holds for the intervals that start when a fishing vessel stops being moored at a $\mt{portA}$, next the vessel is underway and during that period it passes through a fishing area with id  $\mt{AreaID}$ and end when the fishing vessel starts being moored at $\mt{portB}$.
\section{Executable semantics}
\label{sec:exsemantics}
Complex Event Processing refers to the processing of  one or multiple streams of atomic low level entities  by continuously applying temporal queries and producing a stream of time associated complex event detections. In our case, the input stream comprises time associated temporal phenomena and the detection stream comprises the detections of the user defined temporal phenomena. Below, we provide the executable semantics that describe the recognition mechanics for stream processing.

\subsection{Stream processing}

A stream is an arbitrary long sequence of time associated low level entities  i.e., atomic predicates of $\Phi^{\bigcdot}, \Phi^-$ and $\Phi^=$. A stream $\sigma$ that contains time associated atomic formulae `occurring' from the start of time $t_0$ up to a time $t_\sigma$ can be expressed by the model $\mathcal{M}_\sigma=\langle T_\sigma, I_\sigma, <, V^\cdot, V^-, V^=\rangle$, where $T_\sigma=[t_0,t_\sigma]\subset T$ and $I_\sigma=\{ [ts,te] :ts<te\ \text{and}\ ts,te \in T_\sigma\}\cup\{[ts,\infty): ts \in T_\sigma\}$  and as in Section \ref{sec:semantics} $V^.:\Phi_o^.\rightarrow 2^{T_\sigma}$,  $V^-:\Phi_o^-\rightarrow 2^{I_\sigma}$, $V^-:\Phi_o^=\rightarrow 2^{I_\sigma}$ are valuations. 
Note that in the case of streams, a valuation of an atomic formula provides the ordered set of instants or intervals that the atomic formula is associated with. Sets of instants are ordered via the less than relation, while interval sets are totally ordered, in other words the ordering is applied first on the starts of intervals then the ends. This ordering allows, as discussed later, for efficient computations of the instants or the intervals at which formulae hold.

Similar to other approaches~\cite{Artikis_2015,Mileo_2013}, for efficiency reasons, we choose to use a sliding window approach, whereby recognition of user-defined temporal phenomena happens with temporal queries on a dynamically updated working memory $\mt{WM}$ that is specified by a temporal window $\omega$. A temporal window $\omega$ is a finite sub-sequence of a stream $\sigma$, that contains low level entities with associated temporal information that falls within a time period $\omega=(t_i,t_j], t_i<t_j$ with $t_i, t_j \in T$. We denote the number of time instants included in a temporal window $\omega$, i.e., its size, as $|\omega|=t_j-t_i$. 
\subsubsection{Evaluation order of temporal phenomena}
The evaluation of the temporal phenomena must be performed in an order that does not allow unmet dependencies during processing.
Recall that cyclic dependencies are forbidden (see Section~\ref{sec:grammar}), therefore, the dependencies between a set of temporal phenomena can be represented by a directed acyclic graph $G_d=\{E,D\}$ where $E$ is the set of input and user defined phenomena, and $D\subseteq E\times E$ is the set of dependencies.  Consequently, the processing order of a set of temporal phenomena relies on finding a valid evaluation order, that is a numbering $n:E\rightarrow\mathbb{N}$ of the phenomena so that $n(A) < n(B) \Rightarrow  (A,B) \not\in D$, meaning that $A$ will be evaluated before $B$, and that $A$ does not depend on $B$.  A numbering that satisfies these constraints can be acquired by applying topological sort on $G_d$. Phenomena in level 0---these are the input phenomena---of the topological order, have no dependencies, while phenomena in higher levels have at least one direct dependency to phenomena of the previous level, or more to phenomena of lower levels.  
\subsubsection{Sliding window mechanics}
In order to deal efficiently with the input load, we adopt a sliding temporal window approach. With this approach, user defined temporal phenomena are detected using  information inside a working memory set $\mt{WM}$ that contains: time associated phenomena of the input stream that take place within the current temporal window $\omega$, time associated phenomena of previous windows that have been classified as non-redundant and finally the detections of the current query.

Recognition occurs at equally distanced times $t_q \in T$, where $t_{q+1}-t_q=s$ is the sliding step. At each $t_q$, the instants and the intervals at which temporal phenomena are true or hold are computed according to the evaluation order specified by the topological sorting of the $G_d$ graph and the current working memory $\mt{WM}$. Algorithm~\ref{alg:querying} describes the recognition process at a query time $t_q$ with working memory $\mt{WM}$, window size $|\omega|$ and sliding step set to $s$. The instants and intervals at which user defined temporal phenomena are true or hold are computed for each level using `process/2' and are asserted in the current working memory; this ensures that they will be computed only once per recognition query (lines 2-10). During the processing of the user defined temporal phenomena some instants or intervals of the $\mt{WM}$ are classified as non-redundant---we will discuss this step shortly. When all the temporal phenomena levels have been processed, the $\mt{WM}$ is updated (lines 11-12). During this step, all redundant information that falls outside the next window, i.e., within $[t_0,t_q-\omega+s]$, will be removed from the working memory $\mt{WM}$ and the processing of the next query at $t_q+s$ will be able to commence. 
\begin{algorithm}[t]
\small
\KwIn{$\mathit{WM}$; Levels; $|\omega|$; $s$; $t_q$}
allNonRedundant=[]\;
\For{\textnormal{level} \KwInfl \textnormal{[1,Levels]}}{
     phenomenaDefs = phenomenaOfLevel(level)\;
	\For{ \textnormal{phenomenon} \KwInfl \textnormal{phenomenaDefs}}{
		intervals, instants, nonRedundant = process(phenomenon, $\mathit{WM}$)\;
		\eIf{ \textnormal{type(phenomenon==`event')}}{
		$\mt{WM}$.assert(detections(phenomenon, instants))\;}
		{$\mt{WM}$.assert(detections(phenomenon, intervals))\;}
		allNonRedundant.add(nonRedundant)
		}
}
\For{ \textnormal{element} \KwInfl $\mt{WM}$.\textnormal{period(}$t_0$,$t_q-\omega+s$\textnormal{)}}{
\lIf{\textnormal{element} \KwNot \KwInfl \textnormal{allNonRedundant}}{WM.remove(element)}
}
\caption{Querying process for working memory $\mathit{WM}$.}
\label{alg:querying}
\end{algorithm}

\begin{figure}[t]
\centering
\resizebox{.99\textwidth}{!}{%
\begin{tikzpicture}
[x=0.75pt,y=0.75pt,yscale=-1,xscale=1]
\scriptsize
\draw    (70,271) -- (420,271) ;

\draw [shift={(430,270.7)}, rotate = 180.09] [color={rgb, 255:red, 0; green, 0; blue, 0 }  ][line width=0.75]    (10.93,-3.29) .. controls (6.95,-1.4) and (3.31,-0.3) .. (0,0) .. controls (3.31,0.3) and (6.95,1.4) .. (10.93,3.29)   ;


\draw (80,170) -- (300,170);
\draw (80,173) -- (80,167);
\draw (300,173) -- (300,167);
\draw (180,160) node [anchor=north west][inner sep=0.75pt]   [align=left] {$\omega$};
\draw  [dash pattern={on 0.84pt off 2.51pt}]  (80,175) -- (80,280) ;
\draw  [dash pattern={on 0.84pt off 2.51pt}]  (300,177) -- (300,280) ;

\draw (260,177) -- (405,177);
\draw [dash pattern={on 0.84pt off 2.51pt}] (405,177) -- (420,177);

\draw (260,181) -- (260,174);
\draw (342,164) node [anchor=north west][inner sep=0.75pt]   [align=left] {$\omega'$};
\draw  [dash pattern={on 0.84pt off 2.51pt}]  (260,177) -- (260,280) ;


\draw [line width=1, densely dotted]    (100,184.12) -- (120,184.12) ;
\draw [line width=1]    (100,183.12) -- (120,183.12) ;
\draw  [dash pattern={on 0.84pt off 2.51pt}]  (100,184.12) -- (100,271);
\draw  [dash pattern={on 0.84pt off 2.51pt}]  (120,184.12) -- (120,271) ;

\draw [line width=1]    (220,198.12) -- (240,198.12) ;
\draw  [dash pattern={on 0.84pt off 2.51pt}]  (220,198.12) -- (220,271);
\draw  [dash pattern={on 0.84pt off 2.51pt}]  (240,198.12) -- (240,271) ;

\draw [line width=0.3]    (380,197.1) -- (400,197.1) ;
\draw [line width=0.3]    (380,198.9) -- (400,198.9) ;
\draw  [dash pattern={on 0.84pt off 2.51pt}]  (380,198.12) -- (380,271);
\draw  [dash pattern={on 0.84pt off 2.51pt}]  (400,198.12) -- (400,271) ;

\draw [line width=1, densely dotted]    (140,214.12) -- (200,214.12) ;
\draw [line width=1]    (140,213.12) -- (200,213.12) ;
\draw  [dash pattern={on 0.84pt off 2.51pt}]  (140,214.12) -- (140,271);
\draw  [dash pattern={on 0.84pt off 2.51pt}]  (200,214.12) -- (200,271) ;

\draw [line width=1]    (280,213.12) -- (300,213.12) ; 
\draw [line width=0.3]    (300,212.1) -- (360,212.1) ;
\draw [line width=0.3]    (300,213.9) -- (360,213.9) ;

\draw  [dash pattern={on 0.84pt off 2.51pt}]  (280,214.12) -- (280,271);
\draw  [dash pattern={on 0.84pt off 2.51pt}]  (360,214.12) -- (360,271) ;

\draw [line width=1, densely dotted]    (160,229.12) -- (180,229.12) ;
\draw [line width=1]    (160,228.12) -- (180,228.12) ; ;
\draw  [dash pattern={on 0.84pt off 2.51pt}]  (160,228.12) -- (160,271);
\draw  [dash pattern={on 0.84pt off 2.51pt}]  (180,228.12) -- (180,271) ;

\draw [line width=0.3]    (320,227.1) -- (340,227.1) ; 
\draw [line width=0.3]    (320,228.9) -- (340,228.9) ; 
\draw  [dash pattern={on 0.84pt off 2.51pt}]  (320,228.12) -- (320,271);
\draw  [dash pattern={on 0.84pt off 2.51pt}]  (340,228.12) -- (340,271) ;

\draw [line width=1, densely dotted]    (120,244.12) -- (220,244.12) ;
\draw [line width=1]    (120,243.12) -- (220,243.12) ;
\draw [line width=1,color=gray!40]    (240,243.12) -- (310,243.12) ;
\draw [line width=1, densely dotted,color=gray!40]    (310,243.12) -- (405,243.12) ;

\draw [line width=0.3]    (240,257.1) -- (380,257.1) ;
\draw [line width=0.3]    (240,258.9) -- (380,258.9) ;
\draw [line width=1,color=gray!40]    (240,255.12) -- (310,255.12) ;
\draw [line width=1, densely dotted,color=gray!40]    (310,255.12) -- (405,255.12) ;

\draw (430,257) node [anchor=north west][inner sep=0.75pt]   [align=left] {$\mathit{T}$};
\draw (64,281.5) node [anchor=north west][inner sep=0.75pt]   [align=left] {\scriptsize $\mathit{t_{q}-|\omega|}$};
\draw (94,272) node [anchor=north west][inner sep=0.75pt]   [align=left] {\scriptsize $\mathit{t_0}$};
\draw (114,272) node [anchor=north west][inner sep=0.75pt]   [align=left] {\scriptsize $\mathit{t_1}$};
\draw (134,272) node [anchor=north west][inner sep=0.75pt]   [align=left] {\scriptsize $\mathit{t_2}$};
\draw (154,272) node [anchor=north west][inner sep=0.75pt]   [align=left] {\scriptsize $\mathit{t_3}$};
\draw (174,272) node [anchor=north west][inner sep=0.75pt]   [align=left] {\scriptsize $\mathit{t_4}$};
\draw (194,272) node [anchor=north west][inner sep=0.75pt]   [align=left] {\scriptsize $\mathit{t_5}$};
\draw (214,272) node [anchor=north west][inner sep=0.75pt]   [align=left] {\scriptsize $\mathit{t_6}$};
\draw (234,272) node [anchor=north west][inner sep=0.75pt]   [align=left] {\scriptsize $\mathit{t_7}$};
\draw (244,281.5) node [anchor=north west][inner sep=0.75pt]   [align=left] {\scriptsize $\mathit{t_{q'}-|\omega|}$};
\draw (274,272) node [anchor=north west][inner sep=0.75pt]   [align=left] {\scriptsize $\mathit{t_9}$};
\draw (294,281.5) node [anchor=north west][inner sep=0.75pt]   [align=left] {\scriptsize $\mathit{t_{q}}$};
\draw (314,272) node [anchor=north west][inner sep=0.75pt]   [align=left] {\scriptsize $\mathit{t_{9}}$};
\draw (334,272) node [anchor=north west][inner sep=0.75pt]   [align=left] {\scriptsize $\mathit{t_{10}}$};
\draw (354,272) node [anchor=north west][inner sep=0.75pt]   [align=left] {\scriptsize $\mathit{t_{11}}$};
\draw (374,272) node [anchor=north west][inner sep=0.75pt]   [align=left] {\scriptsize $\mathit{t_{12}}$};
\draw (394,272) node [anchor=north west][inner sep=0.75pt]   [align=left] {\scriptsize $\mathit{t_{13}}$};


\draw (-25,253.59) node [anchor=north west][inner sep=0.75pt]   [align=left] {$\mt{fishing\_trip(v,b,f,b)}$};
\draw (-25,238.31) node [anchor=north west][inner sep=0.75pt]   [align=left] {$\mt{fishing\_trip(v,a,f,b)}$};
\draw (-25,223.03) node [anchor=north west][inner sep=0.75pt]   [align=left] {$\mt{in\_fishing\_area(v,f)}$};
\draw (16,208) node [anchor=north west][inner sep=0.75pt]   [align=left] {$\mt{underway(v)}$};
\draw (16,193) node [anchor=north west][inner sep=0.75pt]   [align=left] {$\mt{moored(v,b)}$};
\draw (16,178) node [anchor=north west][inner sep=0.75pt]   [align=left] {$\mt{moored(v,a)}$};
\end{tikzpicture}
}
\caption{Example of redundant (dashed underlined) and non redundant intervals, after the transition from window $\omega$ to $\omega'$. Horizontal black bold and hollow lines correspond to evaluations of phenomena with all the required information, while grey lines correspond to evaluations with incomplete information. Bold black lines denote intervals that are computed during $\omega$, while hollow black lines denote intervals that are computed during $\omega'$.}
\label{fig:window-ex}
\end{figure}
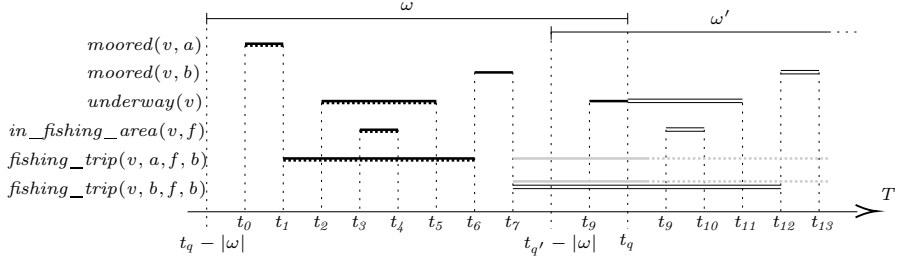

A time associated phenomenon included in $\mathit{WM}$ at $t_q$ is redundant if it holds (a) for an instant or an interval that is is not included or overlaps the next window $(t_q+s-\omega,t_q+s]$ and (b) does not participate in incomplete or complete evaluations of formulae that hold for intervals that overlap the next window.  Consider for example the intervals of Figure~\ref{fig:window-ex}. During the window $\omega=(t_q-|\omega|,t_q]$, the states $\mt{moored(v,a)},\ \mt{moored(v,b)}$ hold for the intervals $[t_0,t_1]$ and $[t_6,t_7]$ respectively, the $\mt{underway(v)}$ state holds for the intervals $\{[t_2,t_5],[t_9,\infty)\}$, the input state $\mt{in\_fishing\_area}$ holds for the interval $[t_3,t_4]$ and finally the dynamic temporal phenomenon $\mt{fishing\_trip(v,a,f,b)}$ holds for the interval $[t_1,t_6]$.  As the window advances, all the intervals that are redundant (dashed underlined) can be discarded since they can no longer contribute in a future query and fall outside the window. However, the interval $[t_6,t_7]$ where $\mt{moored(v,b)}$ holds must be retained, since at window $\omega$ it participates  in the incomplete evaluations of the dynamic temporal phenomena $\mt{fishing\_trip(v,a,f,b)}$ and $\mt{fishing\_trip(v,b,f,b)}$. In other words, it may participate in a detection of $\mt{fishing\_trip(v,a,f,b)}$ or $\mt{fishing\_trip(v,b,f,b)}$ in a future query.

Incomplete evaluations can only occur for dynamic temporal phenomena as per the fact that their detection requires information that sometimes is not yet available. However, this is not the case for evaluations of events and states; since at any given query time and a working memory $\mathit{WM}$ the instants or intervals at which they are true or hold can always be determined. Note though that intervals may be updated if new information allows it. For example, a formula $\phi \rightarrowtail \psi$ can be true at an interval $[ts,\infty)$ at query time $t_q$ while at a query time $t_q'>t_q$ the interval may get updated to $[ts,te]$ if there is a satisfaction of $\psi\wedge\neg\phi$ at $te>t_q$.

The feasibility of an incomplete evaluation of a user defined dynamic temporal phenomenon is determined by propagating the temporal constraints imposed by its definition and by also taking into account the information of the current $\mt{WM}$.  If the evaluation status of a formula $\phi\in\Phi^=$ at query $t_q$ is unknown then all the participating time associated phenomena should be retained for the re-evaluation of $\phi$ in future queries. Consequently, at each query time, in order to label $\mt{WM}$ elements as redundant or not, apart from evaluations of temporal phenomena that can be determined to be \textit{true} or \textit{false}, incomplete evaluations whose validity status is \textit{unknown} must also be computed.

In order to present the mechanism that classifies redundancy we introduce the concept of incomplete intervals. An incomplete interval $i_{inc}=[t,t_\circ]$ is a pair of a known time instant and an unknown value\footnote{We extend the set of all allowed values with $t_\circ$ denoting a time instant that is currently not known but the domain of its possible values is known.}
$\mt{t_{\circ}}$ with a known domain $D_{t_\circ}\subset T$. Incomplete intervals are produced when the evaluation status of a $\phi\in\Phi^=$ formula is \textit{unknown}, while the domains of the unknown instant values are created by propagating the formula constraints. 
Given a working memory $\mt{WM}$ the validity status of a formula  of $\Phi^=$ is \textit{unknown} iff one of the following cases is true:
\begin{itemize}
\item the evaluation status of a formula $\phi\ \plbefore\ \psi$ is \textit{unknown} iff for the last ending interval(s) $\{[ts_1,t],...,[ts_k,t]\}$ where $\phi$ holds  ($\phi \in \Phi^-\cup\Phi^=$) or the last instant $t$ where $\phi$ is true ($\phi \in \Phi^{\bigcdot}$), there not exists an instant $t'$ ($\psi \in \Phi^{\bigcdot}$) or an interval $[t',te']$  ($\psi \in \Phi^-\cup\Phi^=$) with $t'>t$ where $\psi$ is true or holds respectively, in which case the associated incomplete intervals are $[ts_i,t_{\circ}]$, $i\in[1,k]$ when $\phi \in \Phi^-\cup\Phi^=$ or $[t,t_{\circ}]$ when $\phi \in \Phi^{\bigcdot}$ with $D_{t_\circ}=[t+2,\infty)$,
\item the evaluation status of a formula $\phi\ \plcontains\ \psi$ is \textit{unknown} iff $\phi$ holds for an interval $[ts,\infty)$ and there not exists an instant $t$, $t>ts$ ($\psi \in \Phi^{\bigcdot}$) or an interval $[ts',te']$ with $ts<ts'$ ($\psi \in \Phi^-\cup\Phi^=$) where $\psi$ is true or holds respectively, in which case the associated incomplete interval is $[ts,t_\circ]$ with $D_{t_\circ}=(t_q,\infty)$.
\item the evaluation status of a formula $\phi\ \textit{relation}\  \psi$ where \textit{relation} $\in \{\plequals, \plstarts\}$ is \textit{unknown} iff $\phi$ and $\psi$ hold for an interval $[ts,\infty)$ in which case the associated incomplete interval is $[ts,t_\circ]$ with $D_{t_\circ}=(t_q,\infty)$.
\item the evaluation status of a formula $\phi\ \textit{relation}\ \psi$ where $\textit{relation}\in\{\plbefore$, $\plmeets$, $\ploverlaps$, $\plfinishes\}$, $\phi \in \Phi^-\cup\Phi^=$ and $\psi \in \Phi$ is \textit{unknown} iff $\phi$ holds for an interval $[ts,\infty)$, in which case the associated incomplete interval is $[ts,t_\circ]$ with $D_{t_\circ}=[t_q+2,\infty)$.
\item the evaluation status of a formula $\phi\ \mt{relation}\ \psi$ where $\mt{relation}$ is a temporal relation, $\phi,\psi \in \Phi^-\cup\Phi^=$, is \textit{unknown},  iff  the evaluation status of $\phi$ or $\psi$ is $\textit{unknown}$ and the domains of the unknown instants allow a feasible solution to the $\phi\ \mt{relation}\ \psi$ formula's constraints.
\end{itemize} 

At a query time $t_q$, intervals or instants at which a formula $\phi\in \Phi$ is true, ending or occurring before the start of the next window $t_q-\omega+s$, and participating in evaluations with \textit{unknown} validity are classified as non-redundant since their presence will be possibly required in the re-evaluations of the formulae with \textit{unknown} validity in future recognition queries. Moreover, intervals or instants that hold or occur before the start of the next window $t_q-\omega+s$ are also retained if they participate in complete evaluations of formulae that hold for intervals that overlap $t_q-\omega+s$, therefore ensuring that recognised entities that hold on intervals that overlap the next window will also be available for processing in the next recognition query. 
\subsection{Processing of  temporal phenomena}
\label{sec:processing}
The set of time instants or intervals at which input phenomena are true or hold is provided by the input. In the case of user defined phenomena the corresponding sets have to be computed by processing their definitions.
In this section we present the methodology for processing the formulae that comprise temporal phenomena definitions.
\subsubsection{Events}
Events are defined by means of $\phi\ \in\ \Phi^{\bigcdot}$ formulae.
The methodology for computing the set of time instants at which a formula $\phi \in \Phi^{\bigcdot}$ is true is described below.
\begin{itemize}
\item\textbf{Conjunction (Disjunction)} If $\phi$ and $\psi$ are $\Phi^{\bigcdot}$ formulae and are true respectively at the time instants sets $J$ and $K$ $\subseteq\omega$ then their conjunction (disjunction) is described by the set $C=J\cap K$ ($C=J\cup K$).
\item\textbf{Negation} If $\phi$ is a formula of $\Phi^{\bigcdot}$ and $J$ is the set of instants where $\phi$ is true in $\omega$, then $\neg\phi$ is true in the set of instants $C=\omega\setminus J$.

\item\textbf{Start (end)} Start (end) accept a formula $\phi$ of $\Phi^-$ and return the starting (ending) points of the intervals the formula $\phi$ holds.
\end{itemize}
\subsubsection{States}
User-defined states are expressed via the  formulae of $\Phi^-$.  By definition the intervals at which a state may hold are disjoint and maximal; this is because formulae of $\Phi^-$ utilise the temporal operators $\rightarrowtail,\sqcup,\sqcap$ and $\setminus$ which, as defined in Section~\ref{sec:semantics}, will always hold on disjoint maximal intervals. Below we describe the computation of these intervals for each operator.

\textbf{Maximal range} 
Given a working memory $\mt{WM}$ a formula $\phi \rightarrowtail \psi$ for $\phi,\psi \in \Phi$ holds for the intervals that start at the instant $t_s$ where $\phi$ is true and continue to hold indefinitely unless at an instant $t_e>t_s$, $\psi\wedge\neg\phi$ is true. In Appendix~\ref{app:maximalrange} we present the single-scan algorithm that computes the intervals at which a formula $\phi \rightarrowtail \psi$ for $\phi,\psi \in \Phi^{\bigcdot}$ holds.

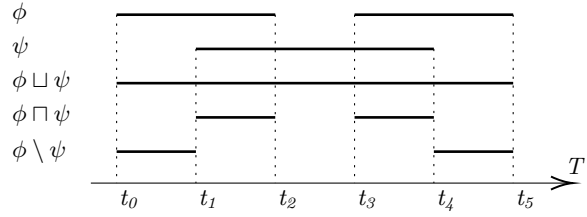
\begin{figure}[t]
\begin{tikzpicture}[x=0.75pt,y=0.75pt,yscale=-1,xscale=1]
\draw    (77,271) -- (320,271) ;

\draw [shift={(320,271)}, rotate = 180.09] [color={rgb, 255:red, 0; green, 0; blue, 0 }  ][line width=0.75]    (10.93,-3.29) .. controls (6.95,-1.4) and (3.31,-0.3) .. (0,0) .. controls (3.31,0.3) and (6.95,1.4) .. (10.93,3.29)   ;

\draw [line width=1]    (90,186) -- (170,186) ;
\draw [line width=1]    (210,186) -- (290,186) ;
\draw  [dash pattern={on 0.84pt off 2.51pt}]  (90,186) -- (90,271) ;
\draw  [dash pattern={on 0.84pt off 2.51pt}]  (170,186) -- (170,271) ;
\draw  [dash pattern={on 0.84pt off 2.51pt}]  (210,186) -- (210,271) ;
\draw  [dash pattern={on 0.84pt off 2.51pt}]  (290,186) -- (290,271) ;

\draw [line width=1]    (130,203.28) -- (250,203.28) ;
\draw  [dash pattern={on 0.84pt off 2.51pt}]  (130,203.28) -- (130,271) ;
\draw  [dash pattern={on 0.84pt off 2.51pt}]  (250,203.28) -- (250,271) ;

\draw [line width=1]    (90,220.56) -- (290,220.56) ;

\draw [line width=1]    (130,237.84) -- (170,237.84) ;
\draw [line width=1]    (210,237.84) -- (250,237.84) ;

\draw [line width=1]    (90,255.12) -- (130,255.12) ;
\draw [line width=1]    (250,255.12) -- (290,255.12) ;

\draw (315,257) node [anchor=north west][inner sep=0.75pt]   [align=left] {$\mathit{T}$};
\draw (90,273.5) node [anchor=north west][inner sep=0.75pt]   [align=left] {$\mathit{t_0}$};
\draw (130,273.5) node [anchor=north west][inner sep=0.75pt]   [align=left] {$\mathit{t_1}$};
\draw (170,273.5) node [anchor=north west][inner sep=0.75pt]   [align=left] {$\mathit{t_2}$};
\draw (210,273.5) node [anchor=north west][inner sep=0.75pt]   [align=left] {$\mathit{t_3}$};
\draw (250,273.5) node [anchor=north west][inner sep=0.75pt]   [align=left] {$\mathit{t_4}$};
\draw (290,273.5) node [anchor=north west][inner sep=0.75pt]   [align=left] {$\mathit{t_5}$};


\draw (36,247.59) node [anchor=north west][inner sep=0.75pt]   [align=left] {$\phi \setminus \psi$};
\draw (36,230.31) node [anchor=north west][inner sep=0.75pt]   [align=left] {$\phi \sqcap \psi$};
\draw (36,213.03) node [anchor=north west][inner sep=0.75pt]   [align=left] {$\phi \sqcup \psi$};
\draw (36,195.75) node [anchor=north west][inner sep=0.75pt]   [align=left] {$\psi$};
\draw (36,178.47) node [anchor=north west][inner sep=0.75pt]   [align=left] {$\phi$};
\end{tikzpicture}
\centering
\caption{Example of the resulting intervals for the temporal operators $\sqcup, \sqcap$ and $\setminus$ between fomulae $\phi$ and $\psi$ of $\Phi^-$. }
\label{fig:tempops}
\end{figure}

\textbf{Temporal union, intersection and complement} In plain language, for $\phi$ and $\psi$ formulae of $\Phi^-$ the temporal operators are defined as follows. The temporal union $\phi\sqcup\psi$, holds for the maximal intervals where at least one of $\phi$ and $\psi$ holds.  The temporal intersection of $\phi\sqcap\psi$ formulae holds for the maximal intervals where both formulae  hold together. Finally, the temporal complement $\phi\setminus\psi$ holds for the maximal sub-intervals of the intervals where $\phi$ holds and $\psi$ does not hold. 
Figure~\ref{fig:tempops} illustrates an example of these operations. The computation of the intervals resulting from these operators can be computed efficiently using single-scan, sorting-based or index-based algorithms~\cite{Dohr2018}. Since these algorithms appear widely in related bibliography~\cite{Dohr2018,bohlen1998}, for our setting, we describe only the single-scan algorithm, presented in~\cite{Dohr2018}, for temporal union in  Appendix~\ref{app:union}. 
\subsubsection{Dynamic temporal phenomena}
Dynamic temporal phenomena are defined using formulae of $\Phi^=$ and may hold in non-disjoint intervals. Given a working memory $\mt{WM}$, the intervals at which dynamic temporal phenomena hold can be computed using the declarative semantics. When the formulae participating in a temporal relation are either of $\Phi^{\bigcdot}$ or $\Phi^-$ efficient linear complexity algorithms are possible since the instants or the disjoint set of intervals at which they hold are ordered. In Appendix~\ref{app:before} we present the linear complexity algorithm that computes the intervals where $\phi\ \plbefore\ \psi$ with $\phi,\psi\in\Phi^-$ holds. When the participating formulae involve those of $\Phi^=$, that may hold on non-disjoint intervals, a naive approach that checks all pairs of the participating time entities requires polynomial time. However, with indexes such as range trees, the intervals at which temporal relations with participating formulae of $\Phi^=$ hold can be computed in linearithmic time~\cite{Mao_Eran_Luo_2019}. 

\section{Related work}
\label{sec:related}
There are numerous languages for describing temporal phenomena with different levels of expressiveness. For example, P. Balbiani  et al. in~\cite{balbiani2011} present a two-sorted point-interval temporal logic framework where both instants and intervals can be used in formulae. However, in P. Balbiani's approach, operations such as the maximal range, union, intersection and complement are not available, therefore phenomena that utilise these operators (e.g., the definition of $\mt{moored/2}$ in Section~\ref{sec:maritime}) are not directly, if at all,  expressible in their language. Moreover, neither executable semantics nor  applications of their temporal logic are discussed. A. Ahmed et al. in~\cite{ahmed2011} use point-based metric temporal logic that has limited interval expressivity to specify stream queries for intrusion detection. Kowalski in~\cite{kowalski86} presents the Event Calculus, a logic based formalism that deals with events (instantaneous) and their effects on inertive fluents (durative), i.e., time-varying properties. 

In the field of complex event processing or recognition, G. Cugola et al.  in~\cite{Cugola_Margara_2010} formally present an event specification language, TESLA, that allows the definition of possibly hierarchical complex event patterns that happen on instants of time. In~\cite{Anicic_2012} D. Anicic et al. present ETALIS, a rule based system where complex events are durative and their definition among others may involve sequences of events, negation and some of the Allen's algebra relations. Compared to TESLA and ETALIS our language allows the description of both instantaneous and durative temporal phenomena. Efficient complex event recognition approaches using the Event Calculus among others involve~\cite{Artikis_2015,chittaro1996}. In~\cite{Artikis_2015} A. Artikis et. al. present the Event Calculus for Run-Time reasoning. Their approach allows the definition and processing of possibly hierarchical events and fluents that happen on instants or hold on intervals. Compared to the Event Calculi approaches of~\cite{Artikis_2015,chittaro1996} our language allows the expression of temporal relations between durative and instantaneous phenomena  that may hold, additionally, in non-disjoint intervals.

\section{Summary \& Future Directions}
\label{sec:summary}
We formally presented a language for the representation of temporal phenomena, with declarative and operational semantics. Our language allows the representation of possibly hierarchical instantaneous and durative temporal phenomena. Definitions of temporal phenomena may involve the standard logical connectives, temporal operators of maximal range, union, intersection and complement and the seven basic temporal relations of Allen's algebra relations of intervals. An open-source Prolog implementation of a Complex Event Processing system called `Phenesthe' utilising the language and the operational semantics presented in this paper is available in~\cite{phenesthe2021}.

Our future directions involve a scalability study  of `Phenesthe' and a demonstration of its efficiency using real world maritime data. Moreover, we aim to evaluate theoretically the expressiveness of our language and compare it with existing ones.  Finally, we aim to integrate temporal stream processing with process mining techniques for the discovery of dynamic temporal phenomena.

\bibliography{bibliography}

\appendix
\section{Algorithms}
\subsection{Maximal range}
\label{app:maximalrange}
Algorithm~\ref{alg:initerm}, describes the computation of the intervals at which a formula $\phi\rightarrowtail\psi$ where $\phi,\psi\in \Phi^{\bigcdot}$ holds. $\mt{ST}$ is a chronologically sorted set comprised of triplets $(t,\phi_{bool},\psi_{bool})$, where $t$ is a time instant at which at least one of $\phi$ or $\psi$ holds, and $\phi_{bool}$, $\psi_{bool}$ are boolean values denoting the truth value of $\phi$ and $\psi$ at the instant $t$. Finally, $\mt{MI}$ is the resulting interval set.

\begin{algorithm}[ht]
\small
\KwIn{$ST$ chronologically sorted set}
\KwResult{$\mt{MI}$ Ordered disjoint interval set}
started=Null\;
\For{$i$ \KwInfl ST}{
	\If{$i.\phi_{bool}$ \KwAnd \textnormal{started==Null}}{
	started=$i.t$}
	\If{$i.\psi_{bool}$ \KwAnd $\neg$ $i.\phi_{bool}$ \KwAnd \textnormal{started$\neq$Null}}{
	MI.append([started,i.t])\;
	started=Null}
}
\If{\textnormal{started $\neq$ Null}}{
	MI.append([started,$\infty$])}
 \caption{Computation of a state's intervals.}
 \label{alg:initerm}
\end{algorithm}
Computation of the intervals where $\phi\rightarrowtail\psi$  holds is achieved by iterating over the instants  where $\phi,\psi$ are true. An interval is added in the $\mt{MI}$ set when the earliest instant at which $\psi\wedge\neg\phi$ is true is processed after the earliest instant at which $\phi$ is true (lines 2-7). An open to infinity interval is added in the $\mt{MI}$ if there isn't an instant where $\psi\wedge\neg\phi$ is true occurring after the earliest instant where $\phi$ is true (8-9). 
\subsection{Temporal union}
\label{app:union}
Consider the formulae $\phi, \psi \in \Phi^-$,  the sets $Int_\phi=\{[ts,te] \in I_\omega: \mathcal{M}_\omega,[ts,te] \models \phi\}$, $Int_\psi=\{[ts,te]\in I_\omega: \mathcal{M}_\omega,[ts,te] \models \psi\}$, then the temporal union $\phi\sqcup\psi$ holds for the intervals computed by the Algorithm~\ref{alg:tunion} as follows. 
\begin{algorithm}[t]
\small
\KwIn{SE ordered starting/ending instants}
\KwResult{UI Ordered disjoint interval set}
starting\_points,ending\_points=0\;
\For{i \KwInfl SE}{
	\If{$i.s$}{
		\lIf{\textnormal{starting\_points}=0}{$ts=i.t$}
		starting\_points+=$i.s$\;
		}
	\If{$i.e$}{
	    ending\_points+=$i.e$\;	
	}
	\If{$s==e$ \KwAnd $s>0$}{
		UI.append([$ts,i.t$])\;
		starting\_points,ending\_points=0\;	
	}
}
\lIf{\textnormal{starting\_points} > \textnormal{ending\_points}}{UI.append([started,$\infty$])}
 \caption{Temporal union.}
 \label{alg:tunion}
\end{algorithm}
`$\mt{SE}$' is a list of triplets $(t,s,e)$ ordered on $t$; $t$ is an instant that is a starting or ending point of an interval in $Int_\phi\cup Int_\psi$ and $s$, $e$ are integer values $\in \{0,1,2\}$ denoting the number of times $t$ is a starting point or an ending point in the intervals of $Int_\phi\cup Int_\psi$. Finally, `UI' contains the intervals where $\phi\sqcup\psi$ holds. `UI' is computed with the use of the two counters `starting\_points' and `ending\_points' corresponding to the number each instant is a starting or ending point respectively. When the two counters become equal an interval that starts from the instant $t$ that first increased `starting\_points' from 0 (lines 4,5) and ends at the instant $t$ that caused `starting\_points==ending\_points' is added in UI (lines 8-10). If by the end of all the iterations, the number of `starting\_points' is greater than the `ending\_points' then an interval open to infinity is appended in UI (line 11). 

\subsection{Temporal relation: \textsf{before}}
\label{app:before}
The intervals at which $\phi\ \plbefore\ \psi$  where $\phi,\psi\in\Phi^-$ holds are computed by Algorithm~\ref{alg:before} in linear time. $\phi_{\mt{ints}}$, $\psi_{\mt{ints}}$ are sets containing the disjoint intervals at which $\phi$ or $\psi$ hold, $\mt{BI}$ is the set containing the computed intervals and $i_{inc}$ is an incomplete interval if available. With the use of the two counters $i,y$, the end of the current interval of $\phi_{\mt{ints}}$ is checked if it is before the current interval of $\psi_{\mt{ints}}$. If this condition is not satisfied the loop will continue until an interval of $\psi_{\mt{ints}}$ is found that satisfies the condition (lines 3-5). When the end of the current interval of $\phi_{\mt{ints}}$ is less than the start of the current interval of $\psi_{\mt{ints}}$ the algorithm continues by checking whether there are any subsequent intervals in $\phi_{\mt{ints}}$ that end later than the start of the current interval of $\psi_{\mt{ints}}$ (lines 6-9). This ensures that the current interval of $\phi_{\mt{ints}}$ is directly before the current interval of $\psi_{\mt{ints}}$. In line 10, the resulting interval is added in $\mt{BI}$. If there are intervals still in $\phi_{\mt{ints}}$  but no intervals of  $\psi_{\mt{ints}}$ remaining incomplete intervals are created according to the rules presented in Section~\ref{sec:exsemantics} (lines 12,15).
\begin{algorithm}[ht]
\small
\KwIn{$\phi_{\mt{ints}}$, $\psi_{\mt{ints}}$ the ordered interval sets where $\phi$ or $\psi$ hold, and $t_q$ the current query time.}
\KwResult{$\mt{BI}$ Ordered  interval set, and $BI_{inc}$ incomplete interval set.}
i,y=0; $i_{inc}$=Null\;
\While{ i  $<|\phi_{\mt{ints}}|$}{
\While{ y  $<|\psi_{\mt{ints}}|$}{
	\lIf{$\phi_{\mt{ints}}$\textnormal{[i].end}  $<\psi_{\mt{ints}}$\textnormal{[y].start}}{\KwBreak}
	y++\;}
\eIf{y  $<|\psi_{\mt{ints}}|$}{
	\While{i+1 $<|\phi_{\mt{ints}}|$}{
			\lIf{$\phi_{\mt{ints}}$\textnormal{[i+1].end}  $<\psi_{\mt{ints}}$\textnormal{[y].start}}{i++}\lElse{\KwBreak}	
	}
	BI.append([$\phi_{\mt{ints}}$[i].start,$\phi_{\mt{ints}}$[i].end])\;
	i++\;
}{
    $BI_{inc}$=create\_incomplete\_intervals($\phi_{\mt{ints}}$,i,$t_q$)\;
   	\KwBreak\;
}
}
 \caption{Computation of the intervals a $\phi\ \plbefore\ \psi$ relation holds for $\phi,\psi\in\Phi^-$.}
 \label{alg:before}
 \end{algorithm}
\end{document}